\documentclass[12pt,titlepage]{article}

\begin{document}

\begin{titlepage}

\title{Observation of Quantum Shock Waves
Created with Ultra Compressed Slow Light Pulses in a Bose-Einstein
Condensate}

\normalsize

\author{
\parbox{6.5in}{
\normalsize Zachary Dutton$^{1,2}$, Michael Budde$^{1,3}$,
Christopher Slowe$^{1,2}$, and Lene Vestergaard Hau$^{1,2,3}$
\newline \newline \\
\normalsize${}^{1}$Rowland Institute for Science, 100 Edwin H.
Land Blvd.,
Cambridge, MA 02142 \\
\normalsize ${}^{2}$Lyman Laboratory, Department of Physics,
Harvard University, Cambridge, MA 02138 \\
 \normalsize${}^{3}$Cruft
Laboratory, Division of Engineering and Applied Sciences, Harvard
University, Cambridge, MA 02138} \\ \\ \\ \\ \\ \\ \\ \\ \\ \\ \\
\\ \\
\hspace{1.5in}
\parbox{4.5in}{
Published online in {\it ScienceExpress} June 28, 2001;\\
\hspace{.5in}10.1126/science.1062527. \\
To appear in print in {\it Science} July 27, 2001. \\
Received May 15, 2001. \\
Accepted June 11, 2001. \\
} }
\date{}

\maketitle

\renewcommand{\baselinestretch}{1.0}
\normalsize

\begin{abstract}

We have used an extension of our slow light technique to provide a
method for inducing small density defects in a Bose-Einstein
condensate.  These sub-resolution, micron-sized defects evolve
into large amplitude sound waves.  We present an experimental
observation and theoretical investigation of the resulting
breakdown of superfluidity.  We observe directly the decay of the
narrow density defects into solitons, the onset of the `snake'
instability, and the subsequent nucleation of vortices.
\end{abstract}

\end{titlepage}
\renewcommand{\baselinestretch}{1.0}
\normalsize

Superfluidity in Bose condensed systems represents conditions
where frictionless flow occurs because it is energetically
impossible to create excitations.  When these conditions are not
satisfied, various excitations develop, and experiments on
superfluid helium, for example, have provided evidence that the
nucleation of vortex rings occurs when ions move through the fluid
faster than a critical speed ({\em 1,2}). Under similar
conditions, shock waves would occur in a normal fluid ({\em 3}).
Such discontinuities are not allowed in a superfluid and instead
topological defects, such as quantized vortices and solitons, are
nucleated when the spatial scale of density variations becomes on
the order of the healing length.  This is the length scale at
which the kinetic energy, associated with spatial variations of
the macroscopic condensate wave function, becomes on the order of
the atom-atom interaction energy ({\em 2,4}).  It is therefore
also the minimum length over which the density of a condensate can
adjust.

Bose-Einstein condensates (BECs) of alkali atoms ({\em 5}) have
provided a system for the study of superfluidity, which is
theoretically more tractable than liquid helium and allows greater
experimental control. An exciting recent development is the
production of solitons and vortices. Experiments have employed
techniques that manipulated the phase of the BEC ({\em 6-9}), or
provided the system with a high angular momentum which makes
vortex formation energetically favorable ({\em 10,11}).  However,
a direct observation of the formation of vortices via the
breakdown of superfluidity has been lacking.  Rather, rapid
heating from 'stirring' with a focused laser beam above a critical
velocity was observed as indirect evidence of this process ({\em
12,13}).

In this context, it is natural to ask what would happen if one
were to impose a sharp density feature in a BEC with a spatial
scale on the order of the healing length. Optical resolution
limits have prevented direct creation of this kind of excitation.
In this paper we present an experimental demonstration of creation
of such defects in sodium Bose-Einstein condensates, using a novel
extension of the method of ultra slow light pulse propagation
({\em 14}) via electromagnetically induced transparency (EIT)
({\em 15,16}).

Our slow-light setup is described in ({\em 14}). By illuminating a
BEC with a `coupling' laser, we create transparency and low group
velocity for a `probe' laser pulse subsequently sent into the
cloud. In a geometry where the coupling and probe laser beams
propagate at right angles, we can control the propagation of the
probe pulse from the side. By introducing a spatial variation of
the coupling field, along the pulse propagation direction, we vary
the group velocity of the probe pulse across the cloud. Here we
accomplish this by blocking half of the coupling beam so it
illuminates only the left hand side ($z<0$) of the condensate,
setting up a light `roadblock'. In this way, we compress the probe
pulse to a small spatial region at the center of the BEC, while
bypassing the usual bandwidth requirements for slow light ({\em
17}). The technique produces a short wavelength excitation by
suddenly removing a narrow disk of the condensate, with the width
of the disk determined by the width of the compressed probe pulse.

We find that this excitation results in short wavelength, large
amplitude sound waves that shed off `gray' solitons ({\em 18-20}),
and we make the connection to the formation of shock waves in
classical fluids.  The `snake' (Kadomtsev-Petviashvili)
instability is predicted to cause solitons to decay into vortices
({\em 21-24}). This has been observed with optical solitons ({\em
25}) and recently the JILA group predicted and observed that
solitons in a BEC decay into vortex rings ({\em 9}). Here we
present a direct observation of the dynamics of the snake
instability in a BEC and the subsequent nucleation of vortices.
The images of the evolution are compared to numerical propagation
of the Gross-Pitaevskii equation in two dimensions.

Details of our Bose-Einstein condensation apparatus can be found
in ({\em 26}).  We use condensates with about 1.5 million sodium
atoms in the state $|1 \rangle \equiv |3S_{1/2},F=1,M_F=-1
\rangle$ and trapped in a 4-Dee magnet.  For the experiments
presented here, the magnetic trap has an oscillator frequency
$\omega_z=(2\pi) 21$ Hz along its symmetry direction and a
frequency $\omega_x = \omega_y = 3.8 \omega_z$ in the transverse
directions. The peak density of the condensates is then $9.1
\times 10^{13}~\mathrm{cm}^{-3}$. The temperature is $T \sim
0.5\,T_c$, where $T_c=300$ nK is the critical temperature for
condensation, and so the vast majority ($\sim 90 \%$) of the atoms
occupy the ground state.

To create slow and spatially localized light pulses, the coupling
beam propagates along the $x$ axis (Fig.~2B), is resonant with the
$D_1$ transition from the unoccupied hyperfine state $|2 \rangle
\equiv |3S_{1/2},F=2,M_F=-2 \rangle$ to the excited level $ |3
\rangle \equiv |3P_{1/2}, F=2, M_F=-2 \rangle$, and has a Rabi
frequency $\Omega_c = (2\pi) 15$ MHz ({\em 27}).  We inject probe
pulses along the $z$ axis, resonant with the
\mbox{$|1\rangle~\rightarrow|3\rangle$} transition and with peak
Rabi frequency $\Omega_p = (2\pi) 2.5$ MHz.  The pulses are
Gaussian shaped with a $1/e$ half-width of $\tau = 1.3\;\mu$s.
With the entire BEC illuminated by the coupling beam, we observe
probe pulse delays of $4\;\mu$s for propagation through the
condensates, corresponding to group velocities of 18 m/s at the
center of the clouds.  A pulse with a temporal half-width $\tau$
is spatially compressed from a length $2 \, c \tau$ in vacuum to
({\em 14,17,28,29})
\begin{equation}
L=2 \tau V_g = 2 \tau \frac{|\Omega_c|^2}{\Gamma f_{13} \sigma_0
n_c}
\end {equation}

\noindent  inside the cloud, where $\Gamma = (2 \pi) 10$ MHz is
the decay rate of state $|3 \rangle$, $n_c$ is the cloud density,
$\sigma_0 = 1.65 \times 10^{-9}\;\mathrm{cm}^2$ is the absorption
cross-section for light resonant with a two-level atom, and
$f_{13}=1/2$ is the oscillator strength of the $|1 \rangle
\rightarrow |3 \rangle$ transition. The atoms are constantly being
driven by the light fields into a dark state, a coherent
superposition of the two hyperfine states $|1 \rangle$ and $ |2
\rangle$ ({\em 15}).  In the dark state, the ratio of the two
population amplitudes is varying in space and time with the
electric field amplitude of the probe pulse as

\begin{equation}
\psi_2 = - \frac{\Omega_p}{\Omega_c}\,\psi_1,
\end {equation}

\noindent where $\psi_1,\;\psi_2$ are the macroscopic condensate
wave functions associated with the two states $|1 \rangle$ and $
|2 \rangle$.

For the parameters listed above, the probe pulse is spatially
compressed from 0.8 km in free space to only 50 $\mu$m at the
center of the cloud, at which point it is completely contained
within the atomic medium.  The corresponding peak density of atoms
in $|2 \rangle$, proportional to $|\psi_2|^2$, is $1/34$ of the
total atom density. The $|1 \rangle$ atoms have a corresponding
density depression.

From Eqs.~1 and 2, it is clear that to minimize the spatial scale
of the density defect, we need to use short pulse widths and low
coupling intensities.  However, for all the frequency components
of the probe pulse to be contained within the transmission window
for propagation through the BEC ({\em 17}), we need a pulse with a
temporal width $\tau$ of at least $2 \sqrt{D} \Gamma /
{\Omega_c}^2 \approx 0.3~\mu$s (here $D \approx 520$ is the
optical density of a condensate for on-resonance two level
transitions) to avoid severe attenuation and distortion.
Furthermore, we see from Eq.~2 that to maximize the amplitude of
the density depression would favor use of a peak Rabi frequency
for the probe of $\Omega_p \sim \Omega_c$. This also severely
distorts the pulse.

Both of these distortion effects accumulate as the pulse
propagates through large optical densities.   This motivated us to
introduce a roadblock in the condensate for a light pulse
approaching from the left hand ($z<0$) side.  By imaging a razor
blade onto the right half of the condensate, we ramp the coupling
beam from full to zero intensity over the course of a $12\;\mu$m
region in the middle of the condensate, determined by the optical
resolution of the imaging system.  In the illuminated region
($z<0$), our bandwidth and weak-probe requirements are well
satisfied and we get undistorted, unattenuated propagation through
the first half of the cloud to the high-density, central region of
the condensate. As the pulse enters the roadblock region of low
coupling intensity, it is slowed down and spatially compressed.
The exact shape and size of the defects which are created with
this method are dependent on when absorption effects become
important.

To accurately model the pulse compression and defect formation, we
account for the dynamics of the slow light pulses, the coupling
field, and the atoms self-consistently. At sufficiently low
temperatures, the dynamics of the two-component condensate can be
modelled with coupled Gross-Pitaevskii (GP) equations ({\em 4,5}).
Here we include terms to account for the resonant two-photon light
coupling between the two components:

\begin{eqnarray}
i \hbar \frac{\partial}{\partial t}\psi_1 & = &
\left(-\frac{\hbar^2 \nabla^2}{2m} + V_1(\mathbf{r}) + U_{11}
|\psi_1|^2 + U_{12}
|\psi_2|^2 \right) \psi_1 \nonumber \\
& & \hspace{2 cm} - i \frac{\mid{\Omega_p}\mid^2}{2 \Gamma} \psi_1
- i \frac{{\Omega_p}^\ast \Omega_c}{2 \Gamma} \psi_2 \nonumber
- i N_c \sigma_e \hbar \frac{k_{2 \gamma}}{2 m} |\psi_2|^2 \psi_1,  \\
i \hbar \frac{\partial}{\partial t}\psi_2 & = &
\left(-\frac{\hbar^2 (\nabla^2 + i \mathbf{k_{2 \gamma}} \cdot
\nabla )}{2m} + V_2(\mathbf{r}) + U_{22} |\psi_2|^2 + U_{12}
|\psi_1|^2
\right) \psi_2 \nonumber \\
& & \hspace{2 cm}- i \frac{\mid{\Omega_c}\mid^2}{2 \Gamma} \psi_2
- i \frac{\Omega_p {\Omega_c}^\ast}{2 \Gamma} \psi_1 - i N_c
\sigma_e \hbar \frac{k_{2 \gamma}}{2 m} |\psi_1|^2 \psi_2.
\end{eqnarray}

\noindent Here $V_1(\mathbf{r}) = \frac{1}{2} m {\omega_z}^2(
\lambda^2 (x^2+y^2) + z^2)$, where $m$ is the mass of the sodium
atoms, and $\lambda=3.8$.   Due to the magnetic moment of atoms in
state $|2 \rangle$, $V_2(\mathbf{r})= - 2 V_1(\mathbf{r})$, and
atoms in this state are repelled from the trap.  The EIT process
involves absorption of a probe photon and stimulated emission of a
coupling photon, leading to a $4.1$ cm/s recoil velocity.  This is
described by a term in the second equation, containing
$\mathbf{k_{2 \gamma}} = \mathbf{k_p - k_c}$, the difference in
wave vectors between the two laser beams.  (Here we use a gauge
where the recoil momentum is transformed away.)  Atom-atom
interactions are characterized by the scattering lengths,
$a_{ij}$, via $U_{ij} = 4 \pi N_c \hbar^2 a_{ij}/m$, where
$a_{11}=2.75\;\mathrm{nm}$, $a_{12}=a_{22} = 1.20\,a_{11}$ ({\em
30}), and $N_c$ is the total number of condensate atoms. To obtain
the light coupling terms in Eq.~3, we have adiabatically
eliminated the excited state amplitude $\psi_3$ ({\em 31}), as the
relaxation from spontaneous emission occurs much faster than the
light coupling and external atomic dynamics driving $\psi_3$. In
our model, atoms in $|3 \rangle$ that spontaneously emit are
assumed to be lost from the condensate, which is why the light
coupling terms are non-Hermitian. Finally, the last term in each
equation accounts for losses due to elastic collisions between
high momentum $|2\rangle$ atoms and the nearly stationary
$|1\rangle$ atoms ($\sigma_e = 8 \pi {a_{12}}^2$) ({\em 32}).

The spatial dynamics of the light fields are described classically
with Maxwell's equations in a slowly varying envelope
approximation, again using adiabatic elimination of $\psi_3$ :

\begin{eqnarray}
\frac{\partial}{\partial z}\Omega_p & = & - \frac{1}{2}f_{13}
\sigma_0 N_c(\Omega_p |\psi_1|^2 +
\Omega_c {\psi_1}^\ast \psi_2), \nonumber \\
\frac{\partial}{\partial x}\Omega_c & = & - \frac{1}{2}f_{23}
\sigma_0 N_c(\Omega_c |\psi_2 \mid^2 + \Omega_p \psi_1
{\psi_2}^\ast).
\end{eqnarray}

\noindent In the region where the coupling beam is illuminating
the BEC ($z<0$), the light coupling terms dominate the atomic
dynamics and solving Eqs.~3 and 4 reduces to Eqs.~1 and 2 above.

We have performed numerical simulations in two dimensions ($x$ and
$z$) to track the behavior of the light fields and the atoms. The
probe and coupling fields were propagated according to Eq.~4 with
a second order Runge-Kutta algorithm ({\em 33}) and the atomic
mean fields were propagated according to Eq.~3, with an
Alternating-Direction Implicit variation of the Crank-Nicolson
algorithm ({\em 33,34}). In this way, Eqs.~3 and 4 were solved
self-consistently ({\em 35}). Profiles of the probe pulse
intensity along $z$, through $x=0$, are shown in Fig.~1A.  As the
pulse runs into the roadblock, a dramatic compression of the probe
pulse's spatial length occurs.  When the probe pulse enters the
low coupling region, the Rabi frequency $|\Omega_p|$ becomes on
the order of $|\Omega_c|$.  So the density of state $|2 \rangle$
atoms, $N_c |\psi_2|^2$, increases in a narrow region, which is
accompanied by a decrease in $N_c |\psi_1|^2$ (Fig.~1B). The
half-width of the defect is $2\;\mu$m. As the compression
develops, absorption/spontaneous emission events eventually start
to remove atoms from the condensate and reduce the probe
intensity.

Experimental results are shown in Fig.~2. Fig.~2A is an in-trap
image of the original condensate of $|1 \rangle$ atoms, Fig.~2B
diagrams the beam geometry, and Fig.~2C shows a series of images
of state $|2 \rangle$ atoms as the pulse propagates into the
roadblock. The corresponding optical density (OD) profiles along
$z$ through $x=0$ are also shown.  The OD is defined to be $-
\mathrm{ln}(I/I_0)$, where $(I/I_0)$ is the transmission
coefficient. All imaging is done with near resonant laser beams
propagating along the $y$-axis, and with a duration of 10 $\mu$s.
There is clearly a build-up of a dense, narrow sample of $|2
\rangle$ atoms at the center of the BEC as the pulse propagates to
the right. Note that the pulse reaches the roadblock at the top
and bottom edges of the cloud before the roadblock is reached at
the center, which is a consequence of the transverse variation in
the density of the BEC, with the largest density along the center
line. After the pulse compression is achieved, we shut off the
coupling beam to avoid heating and phase shifts of the atom cloud
due to extended exposure to the coupling laser, and the subsequent
dynamics of the condensate are observed. (We observed that
exposure to the coupling laser alone, for the exposure times used
to create defects, causes no excitations of the condensates).

In considering the dynamics resulting from this excitation of a
condensate, it should be noted that the roadblock
`instantaneously' removes a spatially selected part of $\psi_1$.
The entire light compression happens in approximately
$15\;\mu\mathrm{s}$.  After the pulse is stopped and the coupling
laser turned off, the $|2 \rangle$ atoms remaining in the
condensate $\psi_2$ have a $4.1$ cm/s recoil and atoms which have
undergone absorption and spontaneous emission events have a
similarly sized but randomly directed recoil.  So the $\psi_2$
component and the other recoiling atoms interact with $\psi_1$ for
less than $0.5$ ms before leaving the region.  Both of these time
scales are short compared with the several millisecond timescale
over which most of the subsequent dynamics of $\psi_1$ occur, as
discussed in the following.

We first considered the one-dimensional dynamics along the $z$
axis.   Snapshots of both condensate components, obtained from
numerical propagation in 1D according to Eqs.~3 and 4, are shown
at various times after the pulse is stopped at the roadblock (and
the coupling laser turned off) (Fig. 3A).  In the linear
hydrodynamic regime, where the density defect has a relative
amplitude $A \ll 1$ and a half-width $\delta \gg \xi$ (here $\xi=
1/\sqrt{8 \pi N_c |\psi_{1}|^2
 a_{11}}$ is the local healing length which is $0.4\;\mu\mathrm{m}$ at the
center of the ground state condensate in our experiment), one
expects to see two density waves propagating in opposite
directions at the local sound speed, $c_s = \sqrt{U_{11} |\psi_1
\mid^2/m}$, as seen experimentally in ({\em 36}). However, for the
parameters used in our experiment, the sound waves are seen to
shed off sharp features propagating at lower velocities.
Examination of the width, speed, and the phase jump across these
features shows that they are gray solitons. Describing the slowly
varying background wave function of the condensate with
$\psi^{(0)}_{1}\!,$ the wave function in the vicinity of a gray
soliton centered at $z_0$ is ({\em 18-20}):

\begin{equation}
\psi_1(z,t) = \psi^{(0)}_{1}(z,t) \left( i \sqrt{1-\beta^2} +
\beta \; \mathrm{tanh}\left(\frac{\beta}{\sqrt{2}~\xi}
(z-z_0)\right) \right).
\end {equation}

\noindent The dimensionless constant $\beta$ characterizes the
`grayness', with $\beta = 1$ corresponding to a stationary soliton
with a $100\%$ density depletion.  With $\beta \ne 1$, the
solitons travel at a fraction of the local sound velocity, $c_s
\sqrt{1-\beta^2}$. As seen in the figure, after a shedding event,
the remaining part of the sound wave continues to propagate at a
reduced amplitude.  Our numerical simulations show that the
solitons eventually reach a point where their central density is
zero and then oscillate back to the other side, in agreement with
the discussions in  ({\em 18,20}).

In Fig.~3A, each of the two sound waves shed off two solitons. By
considering the available free energy created by a defect, one
finds that, when the defect size is somewhat larger than the
healing length and the defect amplitude, $A$, is on the order of
unity, the number of solitons that can be created is approximately
$\sqrt{A} \delta /(2 \xi)$.

One obtains a simple physical estimate of the conditions necessary
for soliton shedding by calculating the difference in sound speed
associated with the difference in atom density between the center
and back edge of the sound wave.  As confirmed by our numerical
simulations, this difference leads to development of a steeper
back edge and an increasingly sharp jump in the phase of the wave
function. This is the analog of shock wave formation from large
amplitude sound waves in a classical fluid ({\em 3}). When the
spatial width of the back edge has decreased to the width of a
soliton with amplitude $\beta = \sqrt{A/2}$ (according to Eq.~5),
such a soliton is shed off the back. It's subsonic speed causes it
to separate from the remaining sound wave.  Furthermore, by
creating defects with sizes on the order of the healing length, we
excite collective modes of the condensate, with wave vectors on
the order of the inverse healing length.  In this regime, the
Bogoliubov dispersion relation is not linear ({\em 4,5}), and
accordingly some of the sound wave will disperse into smaller
ripples, as seen in Fig.~3A.

Considering the evolution of a defect of relative density
amplitude $A$ and half-width $\delta$ in an otherwise homogenous
medium, we estimate that solitons of amplitude $\beta =
\sqrt{A/2}$ will be created after the two resulting sound waves
have propagated a distance

\begin{equation}
z_{sol} = \frac{2 \delta}{A}\left( \frac{1 - \frac{\xi}{2
\delta}}{1 - \frac{{\pi}^2 \xi^2}{\delta^2 A}}\right).
\end {equation}

\noindent This is in agreement with our numerical calculations. We
conclude that the minimum soliton formation length is obtained for
large amplitude defects with a width just a few times the healing
length.  This dictates the defect width picked in the experiments
presented here.  Narrower defects disperse, whereas larger
defects, comparable to the cloud size, couple to collective,
nonlocalized excitations of the condensate.

We explored the soliton formation experimentally by creating
defects in a BEC with the light roadblock.  We controlled the size
of the defects by varying the intensity of the probe pulses, which
had a width $\tau = 1.3~\mu$s.  OD images of state $|1 \rangle$
condensates are shown (Fig.~4) in one particular case. Immediately
after the defect is created, the trap is turned off, and the cloud
evolves and expands for 1 ms and 10 ms, respectively. As seen from
the 1 ms picture, a single deep defect is formed initially, which
results in creation of 5 solitons after 10 ms of condensate
dynamics and expansion. The initial defect created in the trap
could not be resolved with our imaging system, which has a
resolution of 5 $\mu$m.  By varying the probe intensity, we find
that the number of solitons formed scales linearly with the probe
pulse energy, as expected.

To study the stability of the solitons, we first performed 2D
numerical simulations of Eqs.~3 and 4.  Fig.~3B shows density
profiles, $N_c |\psi_1|^2$, obtained for the same parameters as
used in Fig.~1B. Again, the profiles are shown at various times
after the pulse is compressed and stopped. The deepest soliton
(the one closest to the center) is observed to quickly curl and
eventually collapse into a vortex pair.  The wave function
develops a $2 \pi$ phase shift in a small circle around the vortex
cores, which shows that they are singly quantized vortices.  Also,
the core radius is approximately the healing length.  (Upon
collapse, a small sound wave between the two vortices carries away
some of the remaining soliton energy.) This decay can be
understood as resulting from variation in propagation speed along
the transverse soliton front.  As discussed in ({\em 22-25}), a
small deviation will be enhanced by the nonlinearity in the
Gross-Pitaevskii equation, and thus, the soliton collapses about
the deepest (and therefore slowest) point.

Fig. 5 shows experimental images of state $|1 \rangle$
condensates. After the defect is created, the condensate of $|1
\rangle$ atoms is left in the trap for a varying amount of time
(as indicated on the figures).  The trap was then abruptly turned
off and, $15$ ms after release, we imaged a selected slice of the
expanded condensate, with a thickness of 30 $\mu$m along $y$ ({\em
37}).  The release time of $15$ ms is picked large enough that the
condensate structures are resolvable with our imaging system ({\em
38}).  The slice was optically pumped from state $|1 \rangle$ to
the $|3S_{1/2},\,F=2 \rangle$ manifold for $10\;\mu\mathrm{s}$
before it was imaged with absorption imaging by a laser beam
nearly resonant with the transition from $|3S_{1/2},\,F~=~2
\rangle$ to $|3P_{3/2},\,F~=~3 \rangle$. The total pump and
imaging time was small enough that no significant motion due to
photon recoils occurred during the exposure. The slice was
selected at the center of the condensate by placing a slit in the
path of the pump beam.

For the data in Fig.~5A, it is seen that the deepest soliton curls
as it leaves certain sections behind, and at 1.2 ms it has
nucleated vortices.  This is a direct observation of the snake
instability.  In Fig.~5B, at 0.5 ms, the snake instability has
caused a complicated curving structure in one of the solitons, and
vortices are observed after 2.5 ms.  The vortices are seen to
persist for many milliseconds and slowly drift towards the
condensate edge. We observed them even after $30$ ms of trap
dynamics, long enough to study the interaction of vortices with
sound waves reflected off the condensate boundaries.  Preliminary
results, obtained by varying the $y$ position of the imaged
condensate slices, indicate a complicated 3D structure of the
vortices.  In addition, the defect has induced a collective motion
of the condensate whereby atoms, originally in the sides of the
condensate, attempt to fill in the defect.  This leads to a narrow
and dense central region, which then slowly relaxes (Fig. 5B).

We performed the experiment with a variety of Rabi frequencies for
the probe pulses, and saw nucleation of vortices only for the peak
$\Omega_p > (2 \pi) 1.4$ MHz. The free energy of a vortex is
substantially smaller near the border of the condensate where the
density is smaller, so smaller (and thus lower energy) defects
will form vortices very near the condensate edges, seen as
`notches' in Figs.~3B and 5.

In conclusion, we have studied and observed how small wavelength
excitations cause a breakdown of superfluidity in a BEC.  Our
results show how localized defects in a superfluid will quite
generally either disperse into high frequency ripples or end up in
the form of topological defects such as solitons and vortices, and
we have obtained an analytic expression for the transition between
the two regimes.  By varying our experimental parameters, we can
create differently sized and shaped defects, and also control the
number of defects created, allowing studies of a myriad of
effects. Among them are soliton-soliton collisions, more extensive
studies of soliton stability, soliton-sound wave collisions,
vortex-soliton interactions, vortex dynamics, interaction between
vortices, and the interaction between the BEC collective motion
and vortices.

\renewcommand{\baselinestretch}{1.0}
\normalsize

\section*{References}

\begin{enumerate}

\item R.J. Donnelly, {\it Quantized vortices in Helium II}
(Cambridge Univ. Press, Cambridge, 1991).

\item R.M. Bowley, J. Low Temp. Phys. {\bf 87}, 137 (1992).

\item L.D. Landau and E.M. Lifshitz, {\it Fluid Mechanics}
(Pergamon Press, New York, 1959).

\item A.L. Fetter, in {\it Bose-Einstein Condensation in Atomic
Gases, Proceedings of the International School of Physics Enrico
Fermi, Course CXL}, M. Inguscio, S. Stringari, C. Wieman, Eds.
(International Organisations Services B.V., Amsterdam, 1999).

\item For a review, see F. Dalfovo, S. Giorgini,
L.P. Pitaevskii, S. Stringari, Rev. Mod. Phys. {\bf 71}, 463
(1999).

\item M.R. Matthews {\it et al.},
Phys. Rev. Lett. {\bf 83}, 2498 (1999).

\item S. Burger, K. Bongs, S. Dettmer, W. Ertmer,
K. Sengstock, Phys. Rev. Lett. {\bf 83}, 5198 (1999).

\item J. Denschlag {\it et al.}, Science {\bf 287}, 97 (2000).

\item B.P. Anderson {\it et al.}, Phys. Rev. Lett.
{\bf 86}, 2926 (2001).

\item K.W. Madison, F. Chevy, W. Wohlleben,
J.Dalibard, Phys. Rev. Lett. {\bf 84}, 806 (2000).

\item J.R. Abo-Shaeer, C. Raman, J.M. Vogels,
W. Ketterle, Science {\bf 292}, 476 (2001).

\item C. Raman {\it et al.}, Phys.
Rev. Lett. {\bf 83}, 2502 (1999).

\item B. Jackson, J.F. McCann, C.S. Adams, Phys. Rev. A {\bf 61},
051603(R) (2000).

\item L.V. Hau, S.E. Harris, Z. Dutton, C.H.
Behroozi, Nature {\bf 397}, 594 (1999).

\item S.E. Harris, Physics Today {\bf 50}, 36 (1997).

\item M.O. Scully and M.S. Zubairy, {\it Quantum Optics} (Cambridge
Univ. Press, 1997).

\item S.E. Harris and L.V. Hau,
Phys. Rev. Lett. {\bf 82}, 4611 (1999).

\item S.A. Morgan, R.J. Ballagh, K.
Burnett, Phys. Rev. A {\bf 55}, 4338 (1997).

\item W.P. Reinhardt and C.W. Clark,
J. Phys. B {\bf 30}, L785 (1997).

\item Th. Busch and J.R. Anglin, Phys. Rev. Lett.
{\bf 84}, 2298 (2000).

\item B.B. Kadomtsev and V.I. Petviashvili, Sov.
Phys. Dokl. {\bf 15}, 539 (1970).

\item C.A. Jones, S.J. Putterman, P.H. Roberts, J.
Phys. A {\bf 19}, 2991 (1986).

\item C. Josserand and Y. Pomeau, Europhys. Lett. {\bf 30}, 43 (1995).

\item D.L. Feder, M.S. Pindzola, L.A. Collins, B.I. Schneider, C.W.
Clark, Phys. Rev. A {\bf 62}, 053606 (2000).

\item A.V. Mamaev, M. Saffman, A.A. Zozulya, Phys. Rev. Lett.
{\bf 76}, 2262 (1996).

\item L.V. Hau {\it et al.}, Phys. Rev. A {\bf 58}, R54
(1998).

\item We use the definition $\Omega_p
= \mathbf{E}_{p0} \cdot \mathbf{d}_{13}, \Omega_c =
\mathbf{E}_{c0} \cdot \mathbf{d}_{23}$, where
$\mathbf{E}_{p0},\mathbf{E}_{c0}$ are the slowly varying electric
field amplitudes, and $\mathbf{d}_{13},\mathbf{d}_{23}$ are the
electric dipole matrix elements of the transitions.

\item S.E. Harris, J.E. Field, A. Kasapi, Phys.
Rev. A {\bf 46}, R29 (1992).

\item C. Liu, Z. Dutton, C.H. Behroozi, L.V. Hau, Nature
{\bf 409}, 490 (2001).

\item J.P. Burke, C.H. Greene, J.L Bohn, Phys. Rev. Lett. {\bf 81},
3355 (1998).  A Mathematica notebook is available at
http://fermion.colorado.edu/$\sim$chg/Collisions/.

\item J. Javanainen and J. Ruostekoski, Phys. Rev.
A {\bf 52} 3033 (1995).

\item Y.B. Band, M.Tribbenbach, J.P. Burke, P.S.
Julienne, Phys. Rev. Lett. {\bf 84}, 5462 (2000).

\item W. H. Press, S.A. Teukolsky,
W.T. Vetterling, B.P. Flannery, {\it Numerical Recipes in C,
Second Edition} (Cambridge University Press, Cambridge, 1992).

\item S.E. Koonin and D.C. Merideth, {\it Computational Physics},
(Addison-Wesley, Reading, MA, 1990).

\item For propagation of the GP equation in 1D, we typically
used a spatial grid with 4000 points and $dz=0.040\;\mu$m.  In 2D
simulations, we typically used a $750 \times 750$ grid with
$dz=0.21\;\mu$m and $dx=0.057\;\mu$m.  To solve the equations
self-consistently, we kept track of the wave functions at previous
time points and projected forward to second order.  Smaller time
steps and grid spacing were also used to assure convergence of the
results. To mimic the nonlinear interaction strength at the center
of a {\it three} dimensional cloud we put in an effective
condensate radius (calculated with the Thomas-Fermi approximation
({\em 39})) in the dimensions which were not treated dynamically.
In all calculations, the initial condition was the ground state
condensate wave function with all atoms in $|1 \rangle$, obtained
by propagating a Thomas-Fermi wave function in imaginary time.

\item M.R. Andrews, {\it et al.}, Phys. Rev. Lett.
{\bf 79}, 553 (1997).

\item M.R. Andrews {\it et al.}, Science {\bf 275}, 637 (1997).

\item F. Dalfavo and M. Modugno, Phys. Rev. A {\bf
61}, 023605 (2000).

\item G. Baym and C. Pethick, Phys. Rev. Lett. {\bf 76}, 6
(1996).

\item Supplementary material is available at {\em
Science} Online at www.sciencemag.org.

\item This work was supported by the Rowland
Institute for Science, the Defense Advanced Research Projects
Agency, the U.S. Airforce Office of Scientific Research, the U.S.
Army Research Office OSD Multidisciplinary University Research
Initiative Program, the Harvard Materials Research Science and
Engineering Center sponsored by the National Science Foundation,
and by the Carlsberg Foundation, Denmark.  C.S. is supported by a
National Defense  Science and Engineering Grant sponsored by the
Department of Defense.

\end{enumerate}

\pagebreak

\renewcommand{\baselinestretch}{1.0}
\normalsize

\section*{Figure Captions}

\noindent \textbf{Fig.~1.} \textbf{(A)} Compression of a probe
pulse at the light 'roadblock', according to 2D numerical
simulations of Eqs.~3 and 4.  The solid curves indicate probe
intensity profiles along $z$ (at $x=0$), normalized to the peak
input intensity. The snapshots are taken at a sequence of times,
indicated in the figure, where $t=0$ is defined as the time when
the center of the probe pulse enters the BEC from the left. For
reference, the atomic density profile of the original condensate
is plotted (in arbitrary units) as a dashed curve. The gray
shading indicates the relative coupling input intensity as a
function of $z$, with white corresponding to full intensity, and
the darkest shade of gray corresponding to zero.  The spatial turn
off of the coupling field is centered at  $z=0$ and occurs over
$12\;\mu$m, as in the experiment. The number of condensate atoms
is $1.2 \times 10^6$ atoms, the peak density is $6.9 \times
10^{13}\; \mathrm{cm}^{-3}$, and the coupling Rabi frequency is
$\Omega_c=(2 \pi) 8.0$ MHz.  The probe pulse has a peak Rabi
frequency of $\Omega_p=(2 \pi)2.5$ MHz and a $1/e$ half-width of
$\tau = 1.3 \;\mu$s. \textbf{(B)} Creation of a narrow density
defect in a BEC. Density profiles of the two condensate
components, $N_c|\psi_1|^2$ (dashed) and $N_c|\psi_2|^2$ (solid),
along $z$ at $x=0$ for a sequence of times. Note that the $z$
range of the plot is restricted to a narrow region around the
roadblock at the cloud center. The densities are normalized
relative to the peak density of the original condensate indicated
by the red dashed curve.  The other curves correspond to times $1
\;\mu$s (green), $4 \;\mu$s (blue), and $14\; \mu$s (black).  The
width of the probe pulse is $\tau = 4 \;\mu$s and the other
parameters are the same as in (A).  An animated version is
provided in the supplemental material ({\em 40}) as Animation 1.
\\
\\
\noindent \textbf{Fig.~2.} \textbf{(A)} Experimental in-trap OD
image of a typical BEC before illumination by the probe pulse and
coupling field.   The condensate contains $1.5 \times 10^6$ atoms.
The imaging beam was -30 MHz detuned from the $|1 \rangle
\rightarrow |3P_{1/2}, F=2, M_F = -1 \rangle$ transition.
\textbf{(B)} Top view of the beam configuration used to create and
study localized defects in a BEC as discussed in the text.
\textbf{(C)} Build up of state $|2 \rangle$ atoms at the road
block. In-trap OD images (left) show the transfer of atoms from
$|1\rangle$ to $|2 \rangle$ as the probe pulse propagates through
the condensate and runs into the roadblock. The atoms in
$|2\rangle$ were imaged with a laser beam -13 MHz detuned from the
$| 2\rangle \rightarrow |3P_{3/2}, F=3, M_F = -2\rangle$
transition. To allow imaging, the probe pulse propagation was
stopped at various times, indicated in the figure, by switching
the coupling beam off ({\em 29}). The figures on the right show
the corresponding line cuts along the probe propagation direction
through the center of the BEC.  The probe pulse had a Rabi
frequency $\Omega_p=(2 \pi) 2.4$~MHz and the coupling Rabi
frequency was $\Omega_c= (2 \pi) 14.6$~MHz.
\\
\\
\noindent\textbf{Fig.~3.} \textbf{(A)}  Formation of solitons from
a density defect.  The plots show results of a 1D numerical
simulation of Eqs. 3 and 4.  The light roadblock forms a defect
and the subsequent formation of solitons is seen. The defect is
set up with the same parameters for the light fields as in
Fig.~1B. The number of condensate atoms is $N_c=1.2 \times 10^6$
and the peak density is $7.5 \times 10^{13}\;\mathrm{cm}^{-3}$.
The times in the plots indicate the evolution time relative to the
time when the probe pulse stops at the roadblock and the coupling
beam is switched off (at $t=8\;\mu\mathrm{s}$ with $t=0$ as
defined in Fig.~1A). The solid and dashed curves show the
densities of $|1\rangle$ atoms ($N_c | \psi_1|^2$) and $|2\rangle$
atoms ($N_c |\psi_2|^2$).  The phase of $\psi_1$ is shown in each
case with a dotted curve (with an arbitrary constant added for
graphical clarity).  In the first two frames, the $|2\rangle$
atoms quickly leave due to the momentum recoil, leaving a
large-amplitude, narrow defect in $\psi_1$ (Because this is a 1D
simulation, the momentum kick in the $x$ direction is ignored).
\textbf{(B)} The snake instability and the nucleation of vortices.
The plots show the density $N_c |\psi_1|^2$ from a numerical
simulation in 2D, with white corresponding to zero density and
black to the peak density ($6.9 \times
10^{13}\;\mathrm{cm}^{-3}$). The parameters are the same as in
Fig.~1B and the times indicated are relative to the coupling beam
turn-off at $t=21\;\mu\mathrm{s}$.  The solitons curl about their
deepest point, eventually breaking and forming vortex pairs of
opposite circulation (seen first at 3.5 ms). Several vortices are
formed and the last frame shows the vortices slowly moving towards
the edge of the condensate.  At later times, they interact with
sound waves which have reflected off the condensate boundaries.
Animated versions (Animations 2 and 3) are provided in the
supplemental material ({\em 40}).
\\
\\
\noindent\textbf{Fig.~4.} Experimental OD images and line cuts (at
$x=0$) of a localized defect (top) and the resulting formation of
solitons (bottom) in a condensate of $|1 \rangle$ atoms.  The
imaging beam was detuned $-30$ MHz and $-20$ MHz, respectively,
from the $|\,3S_{1/2},\;F=2\rangle \rightarrow |\,3P_{3/2},
F=3\rangle$ transition. Prior to imaging, the atoms were optically
pumped to $|\,3S_{1/2}, F=2\rangle$ for $10\;\mu\mathrm{s}$. The
probe pulse had a peak Rabi frequency $\Omega_p = (2 \pi) 2.4$
MHz. The coupling laser had a Rabi frequency of $\Omega_c = (2
\pi) 14.9$ MHz, was turned on $6\;\mu\mathrm{s}$ before the probe
pulse maximum, and had a duration of $18\;\mu\mathrm{s}$.
\\
\\
\noindent\textbf{Fig.~5.} Experimental OD images of a $|1\rangle$
condensate, showing development of the snake instability and the
nucleation of vortices.  In each case, the BEC was allowed to
evolve in the trap for a variable amount of time after defect
creation. \textbf{(A)} The deepest soliton (nearest the condensate
center) is observed to curl due to the snake instability and
eventually break, nucleating vortices at 1.2 ms. Defects were
produced in BECs with $1.9\times 10^6$ atoms by probe pulses with
a peak $\Omega_p = (2 \pi) 2.4$ MHz, and a coupling laser with
$\Omega_c = (2 \pi)14.6$ MHz.  The imaging beam was $-5$ MHz
detuned from the $|\,3S_{1/2},\,F=2\rangle \rightarrow
|\,3P_{3/2},\,F=3\rangle$ transition. \textbf{(B)} The snake
instability and behavior of vortices at later times.  The
parameters in this series are the same as in (A), except that the
peak $\Omega_p = (2\pi) 2.0$ MHz, the number of atoms in the BECs
was $1.4 \times 10^6\!,$ and the pictures were taken with the
imaging beam on resonance.

\pagebreak

\renewcommand{\baselinestretch}{1.0}
\normalsize

\section*{Animation Captions}

\noindent\textbf{Animation 1.} An animation, based on 2D numerical
calculations, showing creation of a narrow density defect in a BEC
by the light roadblock.  The parameters and conventions are the
same as in Fig.~1B.  Successive frames are spaced by $1~\mu$s. The
solid curve shows the build-up of $|2 \rangle$ atoms as the probe
pulse runs into the roadblock, and the dashed curve shows the
corresponding depletion of the density of the condensate of $|1
\rangle$ atoms.

\noindent\textbf{Animation 2.} Animation of 40 ms of BEC dynamics
based on 1D numerical simulations.  Parameters and conventions are
the same as in Fig.~3A, but the phase is not plotted here.
Successive frames are spaced by 0.25 ms.  The animation shows a
narrow density defect in the $|1 \rangle$ condensate decaying into
four solitons due to the steepening of the back edge of the sound
waves.   The high frequency ripples are due to the nonlinear part
of the Bogoliubov dispersion curve.  When the solitons reach a
point where their amplitude, $\beta$, equals unity they turn
around. Upon reaching the center of the condensate, they pass
through each other unaffected.

\noindent\textbf{Animation 3.} Animation showing 30 ms of dynamics
of the state $|1 \rangle$ condensate, based on 2D numerical
simulations.  Parameters and conventions are the same as in
Fig.~3B.  (The plot range of each frame is $96.8~ \mu$m$~\times~
31.2~\mu$m). Successive frames are spaced by 0.4 ms. The narrow
density defect in the condensate decays into several solitons and
the deepest solitons decay, via the snake instability, into
vortices and release their remaining energy as sound waves.  The
vortices drift slowly, while some of the sound waves reflect off
the condensate boundaries and subsequently interact with the
vortices.

\end{document}